\documentclass{article}
\usepackage{spconf,amsmath,graphicx,color,bm}

\title{The intrinsic value of HFO features as a\\biomarker of epileptic activity.}
\twoauthors
  {Stephen V. Gliske,
   William C. Stacey\sthanks{These authors were partially supported by the Doris Duke Foundation and the US National Institute of Health (NIH) grant 5-K08-NS069783-05.  SG was additionally supported by an NIH Big Data to Knowledge Mentored Training Grant, 1-K01-ES026839-01A1.}}
	{University of Michigan\\
         Department of Neurology\\
         Ann Arbor, Michigan, U.S.A.}
  {Kevin R. Moon, Alfred O. Hero III\sthanks{These authors were partially supported by the US National Science Foundation (NSF) under grant CCF-1217880 and a NSF Graduate Research Fellowship to KM under Grant No. F031543.}}
	{University of Michigan\\
	Electrical Engineering and Computer Science\\
	Ann Arbor, Michigan, U.S.A.}

\begin{document}
\ninept
\maketitle
\begin{abstract}
High frequency oscillations (HFOs) are a promising biomarker of
epileptic brain tissue and activity.  HFOs additionally serve as a
prototypical example of challenges in the analysis of discrete events
in high-temporal resolution, intracranial EEG data.  Two primary
challenges are 1) dimensionality reduction, and 2) assessing
feasibility of classification.  Dimensionality reduction assumes that
the data lie on a manifold with dimension less than that of the
features space.  However, previous HFO analysis have assumed a linear
manifold, global across time, space (i.e. recording
electrode/channel), and individual patients.  Instead, we assess both
a) whether linear methods are appropriate and b) the consistency of
the manifold across time, space, and patients.  We also estimate
bounds on the Bayes classification error to quantify the distinction
between two classes of HFOs (those occurring during seizures and those
occurring due to other processes).  This analysis provides the
foundation for future clinical use of HFO features and guides the
analysis for other discrete events, such as individual action
potentials or multi-unit activity.
\end{abstract}
\section{Introduction}
\label{sec:intro}

\begin{figure}[htb]
\centering
\centerline{\includegraphics[width=3.5in]{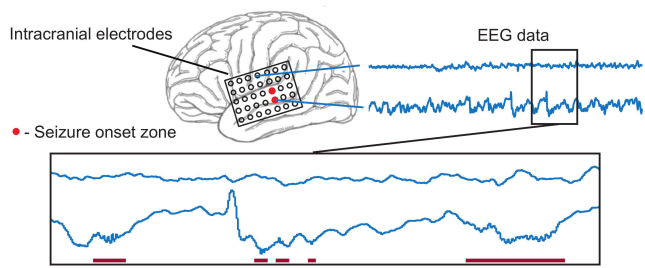}}
\caption{Diagram relating the recorded data with the implanted
  electrodes.  A $5 \times 7$ grid of electrodes placed over a region
  of cortex.  Each channel produces a separate time series of data,
  with some channels being identified by clinicians as seizure
  onset zone (SOZ).  HFO detections are marked by solid magenta lines
  under the EEG trace.}
\label{fig:diagram}
\end{figure}

About one third of epilepsy patients fail to obtain seizure control
with available pharmaceuticals.  One of the few options for these
refractory patients is resective surgery---removing the portion of the
brain thought to be causing the seizures.  This region is denoted the
seizure onset zone (SOZ).  In some cases, determining the SOZ involves
a highly invasive surgery to place electrodes on the brain's surface,
followed by one to two weeks of recording and monitoring.  A second
invasive surgery is performed if the SOZ can be identified and safely
resected.  A schematic relating the implanted electrodes with the
recorded data is shown in Fig~\ref{fig:diagram}.

\begin{figure}[htb]
\centering
\centerline{\includegraphics[width=3.5in]{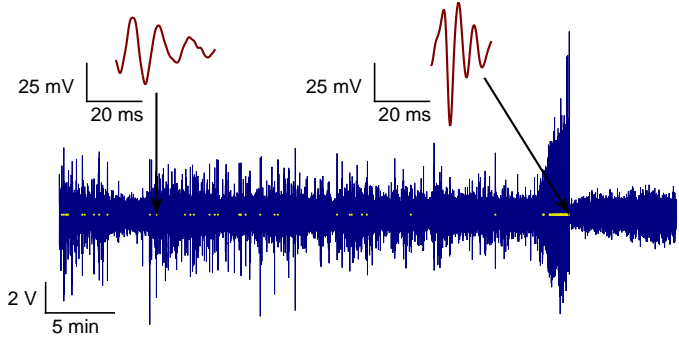}}
\caption{Example HFO detections within 45 min of one channel
  of intracranial EEG data.  A seizure occurs at about
  35 minutes.  HFO detections (72 interictal and preictal, 32 ictal)
  are shown as small yellow dots.  Two HFOs are
  also shown using a \emph{much} smaller scale.}
\label{fig:hfo_ex}
\end{figure}

A proposed biomarker to improve the localization of the SOZ are high
frequency oscillations (HFOs)~\cite{HFO_bkg_1,HFO_bkg_2}.  HFOs are
high frequency (about 80--300 Hz), short ($<50$~ms), rare events
occurring in intracranial EEG recorded at sampling rates of several
kHz.  Example HFO detections and a recorded seizure are shown in Fig~\ref{fig:hfo_ex}.

Much of the published research on HFOs uses human identified HFOs in
short (10 to 20 minute) recordings~\cite{HFO_human_1,HFO_human_2}.
These results have shown that a high HFO occurrence rate is correlated
with the SOZ.  However, recent work is moving towards automated
identification and analysis of HFOs in long term, high resolution
data, which requires advanced computational and statistical
techniques~\cite{qHFO}.  Quality recordings may span 7-14 days, with
over 100,000 HFO detections in several terabytes of data.  Thus, the
next advances are expected to come through big data analysis of HFO
features, utilizing millions of recorded HFOs across as many patients
as possible~\cite{Worrell}.

Relatively few research groups have analyzed HFO features in detail.
The most advanced analysis computed six features of about 300,000 HFOs
in nine patients and two controls, and utilized a global PCA across
all channels followed by $k$-means
clustering~\cite{Blanco_2010,Blanco_2011,Pearce_2013}.  The authors
implicitly assumed that the distribution of these HFO features lies on
a linear manifold in feature space, and that the manifold is
consistent across time and space (i.e., recording electrode).

The other most advanced analysis compared HFOs produced in the motor
cortex via movement versus HFOs occurring in the SOZ,
utilizing three features and a support vector machine (SVM)
classifier~\cite{finger_hfo}.  Some differences were noted, but a
more general analysis that addresses the degree to which HFOs produced
by pathological activity or networks (denoted pathological HFOs or
pHFOs) and HFOs produced by normal, physiological activity (denoted
normal HFOs or nHFOs) are observably different has not been performed.

The goals of this work are to test the implicit assumptions previously
used in HFO feature analysis.  Specifically, the goals are to
1)~assess the type of manifold on which HFO features lie, and 2)
assess how discernible pHFOs are from nHFOs, based on their
feature-space distributions.

The general outline is as follows.  To assess the linearity of the
HFO-feature manifold, a non-linear, local estimate of intrinsic
dimension~\cite{kNN_idim,costa2006dim} is compared with a linear
global estimate (PCA) applied to local subsets. This approach is
similar to that in~\cite{Moon_2015_sunspot}.  The corresponding
reduced dimension subspaces are individually compared across time and
space using a modified Grassmann distance, building off the work
in~\cite{gen_Grassmann}.  We additionally use a greedy Fisher LDA
algorithm (similar to the greedy LDA in~\cite{greedy_LDA}) to identify
a basis that maximally separates pHFOs from nHFOs.  Unsupervised
clustering of the subspaces is then compared with channel groups based
on clinical markings of the SOZ and physical groupings of the
electrodes.  Lastly, we assess the discernibility of nHFOs and pHFOs
by estimating bounds on the Bayes Error using the
Henze-Penrose divergence~(HPD)~\cite{Moon_2015_meta,HP_div}.

\section{Patient population and data}

EEG data from adult patients with refractory epilepsy who underwent
intracranial EEG monitoring were selected from the IEEG
Portal~\cite{IEEG} and from the University of Michigan. All patient
data was included which met the following criteria: sampling rate of
at least 5~kHz, recording time greater than one hour, data recorded
with traditional intracranial electrodes, and available meta-data
regarding seizure times and the resected volume or SOZ.  This yielded
17 patients, (nine IEEG portal, eight U. of M.).  All data were
acquired with approval of local institutional review boards (IRB), and
all patients consented to share their de-identified data.  Of these 17
patients, 13 had recorded seizures, nine were known to have resection
and obtained seizure-freedom (ILAE class I), with eight patients in
common between these two categories.  Because these surgeries are
relatively rare, this patient population size is moderately large for
analysis of intracranial EEG data.


HFOs were detected using the qHFO algorithm~\cite{qHFO}, resulting in
over 1.6 million HFOs in nearly 100,000
channel-hours of 5 kHz data.  Each HFO was band pass filtered between
80 to 500 Hz using an elliptical filter and 33 features were computed,
including duration, peak power, mean of the Teager-Kaiser
energy~\cite{TeagerKaiser}, and various spectral properties.  Ictal is
defined as during seizures, with seizures assumed to be five minutes
long if the length was not specified in available meta-data.
Interictal is defined as at least 30 minutes from the start or end of
a seizure, based on~\cite{Pearce_2013}.

\section{Dimensionality reduction}
\label{sec:dim_red}

\subsection{Consistency of local, non-linear intrinsic dimension}
\label{sec:knn_idim}

To assess both the linearity and local versus global nature of the HFO
feature manifold, the non-linear intrinsic dimension was computed via
the $k$-nearest neighbor (NN) based estimator in~\cite{kNN_idim}.
This estimator estimates the intrinsic dimension by exploiting its
relationship to the total edge length of the $k$-NN graph.  This
nonlinear estimator provides a $\emph{local}$ estimate of intrinsic
dimension which enables us to identify local variations in data
manifolds.  The result is an estimate of the intrinsic dimension for
each given HFO, which are then averaged to obtain the mean intrinsic
dimension for a given partition of the data.  The
consistency of the manifold is measured by comparing the distribution
of intrinsic dimension across time, space and patients.  The specific
comparisons are: a)~interictal versus ictal times, per channel, b)~time
variation within interictal periods, per channel, and c)~comparisons
between channels, integrated over time.

A variety of methods could be employed to compare the intrinsic
dimension for two disjoint sets of HFOs.  However, the final dimension
selected will be an integer.  Thus, small differences
in the intrinsic dimension between two sets, no matter how
statistically significant, are not meaningfully different if they
mean value for each set round to the same integer.

To compare two sets of intrinsic dimension, we define a distance
measure, $\theta_I$, between collections of integers.  Let the two
sets of integers be $A$ and $B$, and let $n_i$ be the fraction
of elements in $A$ equal to $i$, and $m_i$ be the fraction of elements in $B$ equal to $i$. The probability that two elements in set $A$ are
equal is $\bm n^T \bm n$, and the probability that an element of $A$
is equal to an element of $B$ is $\bm n^T \bm m$.  A measure of
distance between $A$ and $B$ is how likely an element of $A$ is equal
to an element of $B$, normalized by the likelihoods of elements being
equal another element in the same group:
\begin{equation}
\label{eq:theta_I}
  \theta_I = \arccos\left( \frac{\bm n^T \bm m }{ \sqrt{\bm n^T \bm n}\sqrt{\bm m^T \bm m} } \right),
\end{equation}
The value $\theta_I$ is the angle between $\bm n$ and $\bm m$ and
provides an easy interpretation as to the consistency of two different
collections of local intrinsic dimensional.  This quantity is also
know as the angular distance or angular dissimilarity, and is the
inverse cosine of the Ochiai-Barkman
coefficient~\cite{Ochiai,Barkman}.

\begin{figure}[htb]
\centering
\centerline{\includegraphics[width=3.5in]{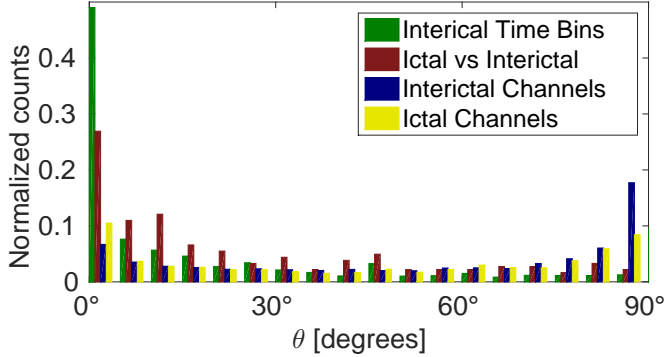}}
\caption{Comparisons of the consistency of the intrinsic dimension
  ($\theta_I$ from Eq.~\ref{eq:theta_I}) for four comparisons, as
  described in the text.  Note, $0^\circ$ implies no difference and
  $90^\circ$ implies maximal difference between the two collections of
  intrinsic dimension being compared.}
\label{fig:theta_I}
\end{figure}

We compute the $\theta_I$-distance for three different comparisons of
HFOs: 1)~a comparison of each pair of 30 minute time windows on a
given channel for a given patient, 2)~a comparison of ictal versus
interictal periods, again on a given channel for a given patient, and
3)~a comparison of different channels in a given patient during
interictal periods.  Interictal-ictal comparisons where one set of
events is less than 50 HFOs are ignored.  Histograms of the
distribution of $\theta_I$ for each type of comparison are shown in
Fig.~\ref{fig:theta_I}.  This figure involves over 5~million
interictal time bin comparisons, 163~ictal versus interictal
comparisons, over 36~thousand interictal channel-channel comparisons
and almost 10~thousand ictal channel-channel comparisons.

The intrinsic dimension is quite consistent across different time
segments during interictal times (strong peak near $0^\circ$), but has
some variance between ictal and interictal times (still peaked near
$0^\circ$ but the peak is wider).  The dimension is less consistent
across channels, with comparisons during ictal times showing small
peaks at both $0^\circ$ and $90^\circ$, and interictal having the
largest peak at $90^\circ$, showing maximal difference.  Thus we see
that the intrinsic dimension varies significantly across channels,
especially for interictal HFOs.

\subsection{Comparison with Global Linear Intrinsic Dimension}

Next we compare the $k$-NN intrinsic dimension estimate of
Section~\ref{sec:knn_idim} with a global, linear method to assess how
linear and/or local the feature manifold is.  The most common global,
linear method of intrinsic dimension estimation and dimensionality
reduction is principle component analysis (PCA), which we perform by
first centering the data and then using singular value decomposition
(SVD).  PCA is performed multiple times for the same divisions of the
HFO data as done for Fig.~\ref{fig:theta_I}.  Subsets of HFOs
with less than 50 events are ignored, as these are deemed insufficient
to estimate the PCA vectors in 33D.  This results in PCA vectors being
computed for 606 of the 1318 channels (12 patients) for interictal
HFOs, 171 channels for ictal HFOs (8 patients), and 163 channels
comparing ictal versus interictal (8 patients).  In patients
that had multiple recording sessions, channels are counted once per each
session.

To compare the $k$-NN (non-linear) and PCA intrinsic dimension per
channel, we 1) select the number of principle components equal to the
non-linear intrinsic dimension estimate, and then 2) report the
fraction of the variance accounted for by that number of principle
components.  This is repeated for all 606 channels (interictal) and
171 channels (ictal).  When using either ictal or interictal HFOs, the
median fraction of variance was 99.8\%, with 99\%-tile of the channels
being above 89.5\% (interictal) and 97.1\% (ictal).  It is likely that
that noise in the data could account for up to 10\% of the variance,
and thus we conclude that the feature manifolds are approximately
linear over the locality of a given channel and ictal state.

\subsection{Comparison between Local Manifolds}

In addition to the earlier comparison of the subspace dimensionality
across channels, the next step is to directly compare the subspaces
selected by the dimensionality reduction.
We use a generalization of distance in the Grassmann space, which
allows comparison of affine subspaces with unequal
dimension~\cite{gen_Grassmann}.  The method augments the principle
angles (defined in~\cite{Hotelling}) with enough additional
angles (all equal to $\pi/2$) to increase the number of angles to the
dimensionality of the larger space.  An additional ``direction
vector'' is also added to account for the affine offset.

We apply this generalization~\cite{gen_Grassmann} to a new modification
of the chordal distance, defined as
\begin{equation}
\label{eq:theta_C}
\theta_C = \arcsin\left( \left( \frac{1}{k} \sum_{i=1}^k \sin^2\theta_i \right)^{1/2} \right),
\end{equation}
for $k$ principle angles $\{ \theta_i \}_{i=1}^k$.  The two modifications are
1)~dividing by $k$, which allows the distance to be independent on the
dimensionality of the spaces being compared, and 2)~converting the
distance measure back to an angle, which is more intuitive.

We then compare the subspaces obtained by PCA dimensionality
reduction, for the same divisions of the data as used for
Fig.~\ref{fig:theta_I}.  However, we now ignore any subsets of less
than 50 HFOs, as these are deemed unreliable for computing the PCA in
33D.  Note this figure still involves over 200~thousand time bin
comparisons, the same 163~ictal versus interictal comparisons, and nearly
3,500 of each type of channel-channel comparisons.

\begin{figure}[htb]
\centering
\centerline{\includegraphics[width=3.5in]{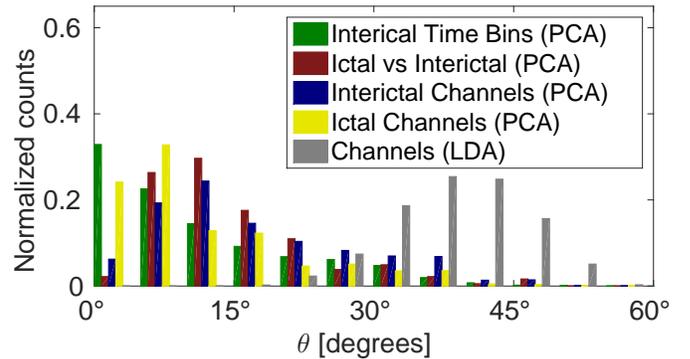}}
\caption{Comparison of the subspaces from the PCA and greedy Fisher
  LDA dimensionality reduction, using $\theta_C$
  (Eq.~\ref{eq:theta_C}), for the same types of comparisons as
  Fig.~\ref{fig:theta_I}.  Again, $0^\circ$ implies maximally similar
  and $90^\circ$ implies maximally different.}
\label{fig:theta_C}
\end{figure}

Results for these comparisons are shown in Fig.~\ref{fig:theta_C}.  We
observe that the PCA subspaces are quite consistent across different
time bins, with the distributions for PCA subspaces all peaking less
than $15^\circ$ and not extending much past $20^\circ$.

\section{Pathological versus Normal HFOs}

\subsection{Dimensionality Reduction: Greedy Fisher LDA}

While the subspaces obtained via PCA represent the variance well, they
are not necessarily the optimal directions for separating the feature
distributions of pHFOs and nHFOs.  An alternate dimensionality
reduction method is used: greedy Fisher's LDA.  In this method,
Fisher's LDA is applied, resulting in a single basis direction.  The
projection of the data in this direction is then subtracted from the
data, and the process is repeated.

Recall that Fisher's LDA uses the sum of the covariance of each group,
rather than the covariance of the pooled groups~\cite{Fisher}.
Letting the mean and covariance of the two groups be denoted $\mu_A$,
$\mu_B$ and $\Sigma_A$, $\Sigma_B$, respectively, the specific direction is given by
\begin{equation}
w \propto \left( \Sigma_A + \Sigma_B \right)^{-1} \left( \bm \mu_A - \bm \mu_B \right).
\end{equation}
Note, the rank of the sum of the covariances will be reduced by one in
each step.  Thus, to invert the matrix, the eigenvalue method is
used, with the inverse of the eigenvalues corresponding to the
removed projections being set to zero.

We compute this basis, comparing ictal versus interictal times, for
the 13 patients with recorded seizures.  We select the number of
basis vectors equal to the mean intrinsic dimension.
%
%
We again compare the basis vectors using Eq.~\ref{eq:theta_C}, with
the results shown in Fig.~\ref{fig:theta_C}.  The LDA subspaces vary
much more across channels than the PCA subspaces, with the $\theta_C$
distribution for LDA being almost fully localized between $30^\circ$
and $50^\circ$.  Note that $45^\circ$ implies that the subspaces
overlap by half.  Thus, the Fisher LDA subspaces show that some
differences in ictal versus interictal HFO features are consistent
between channels, while other differences are not conserved.


\subsection{Bayes Error Estimates of pHFOs versus nHFOs}

Next we quantify how distinct the feature distributions of pHFOs and
nHFOs are, per channel.  This quantification serves as a guide for
future work.  We utilize the Henze-Penrose divergence
(HPD)~\cite{Moon_2015_meta,HP_div} to compute bounds on the
Bayes Error.  Note that small upper bounds imply highly separable
classes, whereas upper bounds near 0.5 imply inseparable classes.  We
estimate the HPD bound using the nonparametric ensemble estimator
derived in~\cite{moon2014ensemble, moon2014multivariate}, which
achieves the parametric convergence rate.

\begin{figure}[htb]
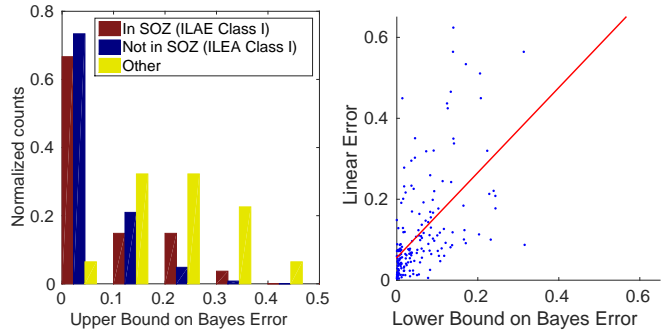

\centering
\centerline{\includegraphics[width=1.7in]{HP_div_par_set_A.eps}
\includegraphics[width=1.7in]{BE_vs_cv_err_par_set_A.eps}}\
\caption{Distribution of upper Bayes Error bound based on
  Henze-Penrose divergences (HPD) between ictal and interictal HFOs
  per channel.  Left panel: upper bounds stratified by channel type.
  Bounds near zero imply high distinction between pHFOs and nHFOs, and
  bounds near 0.5 imply no distinction. Right panel: comparison
  between lower bound estimate and a linear error estimate (see the
  text), including the linear regression fit.}
\label{fig:HPD}
\end{figure}

The left panel of Fig.~\ref{fig:HPD} shows the Bayes Error bound
estimates for the 163 channels (eight patients) with at least 50 HFOs
in each of the ictal and interictal states.  Channels are separated
between SOZ (27~channels) and non-SOZ (124~channels) for patients with
ILEA Class I (the best) surgery outcome, and an ``other'' category
(12~channels), including channels from patients with either worse
surgery outcomes, no surgery, or missing meta-data.

Patients with Both SOZ and non-SOZ channels in ILAE class I patients
(best surgery outcome) have the Bayes Error rate bound to relatively
small values.  However, channels in other patients have a more diffuse
distribution of upper bounds, with the most probable upper bound about
0.25.  It is expected that ictal HFOs are almost entirely pHFOs (on
any channel), that interictal HFOs on non-SOZ channels are
predominately nHFOs, and that interictal HFOs on SOZ channels are a
mixture of both pHFOs and nHFOs.  Thus, it is expected that the Bayes
Error bounds would be lower for non-SOZ channels.  Overall, these
bounds suggest that there is sufficient separation between pHFO and
nHFO feature distributions to allow classification of pHFOs and nHFOs
in most channels.

The right panel of Fig.~\ref{fig:HPD} displays the lower bound versus
a linear error estimate.  The linear estimate was computed with
10-fold cross validation in the Fisher LDA space by using a ``box''
classification boundary, with the threshold in each dimension being
the value for which the receiver operator curve has largest transverse
distance from the diagonal. Linear regression was also performed,
resulting in an offset of 0.06 (0.04--0.08 at 95\% C.L.) and a slope
of 1.05 (0.82--1.28 at 95\% C.L.).  Thus, in aggregate this ``box''
classifier is already relatively close to the bound on the Bayes
Error, though many individual channels are still quite far from the bound.

\section{Clustering Channels based on subspaces}

Given the observed variations in greedy Fisher LDA
subspaces across channels, we seek to compare the natural clustering
of these subspaces with known groupings of channels.  The most
relevant groupings of channels are the groups based on the physical
configuration of the recording electrodes (several grids or strips of
electrodes are implanted for each patient), as well as the clinically
determined SOZ and resected volume for ILAE Class I
patients.

The unsupervised clustering is obtained by first converting the matrix
of modified chordal distances (Eq.~\ref{eq:theta_C}) between channels
per each patient to a metric using the method of dual-rooted-trees
followed by spectral clustering~\cite{DRT_alg}.  This method adapts to
the natural geometry of the data and is competitive with other
algorithms~\cite{DRT_alg}. We select either two groups (as the SOZ and
resected volume clustering is binary) or the number of groups equal to
the number of strips and grids.  The unsupervised clustering results
are compared with these labels using the adjusted Rand index
(ARI)~\cite{ARI}.  Unfortunately, the requirement that there be at
least 50 ictal HFOs reduces the number of channels per patient
significantly, resulting in only a few (4--5) patients having enough
channels to cluster.  However, the ARI never exceeded 0.02 for any of these comparisons in any patients, suggesting that the primary
distinction between channels is not the pathology or the grid/strip
placing, but other effects.


\section{Discussion and Conclusion}

Overall, we observe that the HFO features tend to cluster on linear
manifolds.  Both the subspaces of these manifolds and the local
intrinsic dimension tend to be consistent within interictal periods,
but may change between interictal and ictal periods.  We especially
note significant differences in both the intrinsic dimension and
feature manifolds between different channels within the same patient.
Thus, dimensionality reduction and feature analysis must account for
variations between channels.  The dominant cause of this
inter-channel variation does not appear to be tissue pathology or
grid/strip groups in the recording.  We also observe that pHFOs and
nHFOs are indeed distinct on a large number of channels, suggesting a
strong potential for classifying individual HFOs.

This analysis also demonstrates methods applicable to other discrete
events, including using the $\theta_I$ statistic for comparing local,
intrinsic dimension for collections of events, an affine Grassmann
distance $\theta_C$ for comparing consistency of subspaces, and
estimating bounds on the Bayes Error to assess the feasibility of
low-error classification. Future work will extend this analysis to a
larger patient population, and classify HFOs and/or recording channels
based on HFO features.


\vfill\pagebreak


\bibliographystyle{IEEEbib}
\bibliography{icassp_bib}

\end{document}